\UseRawInputEncoding
\documentclass[onecolumn,superscriptaddress]{revtex4-2}

\usepackage{amsmath,amssymb,amsfonts}
\usepackage{graphicx}
\usepackage{braket}
\usepackage{xcolor}
\usepackage{bm}
\usepackage{booktabs} 

\newcommand{\CP}{\mathbb{CP}}
\newcommand{\R}{\mathbb{R}}
\newcommand{\C}{\mathbb{C}}
\newcommand{\su}{\mathfrak{su}}

\newcommand{\SU}{\mathrm{SU}}

\begin{document}

\title{Quantum-Inspired Unitary Pooling for Multispectral Satellite Image Classification}

\author{Georgios Maragkopoulos}
\email{giorgosmarag@di.uoa.gr}
\affiliation{Department of Informatics and Telecommunications,
National and Kapodistrian University of Athens, Greece}
\affiliation{Eulambia Advanced Technologies Ltd, Ag. Paraskevi, Greece}

\author{Aikaterini Mandilara}
\affiliation{Department of Informatics and Telecommunications,
National and Kapodistrian University of Athens, Greece}
\affiliation{Eulambia Advanced Technologies Ltd, Ag. Paraskevi, Greece}

\author{Ralntion Komini}
\affiliation{Department of Informatics and Telecommunications,
National and Kapodistrian University of Athens, Greece}
\affiliation{Eulambia Advanced Technologies Ltd, Ag. Paraskevi, Greece}

\author{Dimitris Syvridis}
\affiliation{Department of Informatics and Telecommunications,
National and Kapodistrian University of Athens, Greece}
\affiliation{Eulambia Advanced Technologies Ltd, Ag. Paraskevi, Greece}

\begin{abstract}
Multispectral satellite imagery poses significant challenges for deep learning models due to the high dimensionality of spectral data and the presence of structured correlations across channels. Recent work in quantum machine learning suggests that unitary evolutions and Hilbert-space embeddings can introduce useful inductive biases for learning. In this work, we show that several empirical advantages often attributed to quantum feature maps can be more precisely understood as consequences of geometric structure induced by unitary group actions and the associated quotient symmetries.
Motivated by this observation, we introduce a fully classical pooling mechanism that maps latent features to complex projective space via a fixed-reference unitary action. This construction effectively collapses non-identifiable degrees of freedom, leading to a reduction in the dimensionality of the learned representations.
Empirical results on multispectral satellite imagery show that incorporating this quantum-inspired pooling operation into a convolutional neural network improves optimization stability, accelerates convergence, and substantially reduces variance compared to standard pooling baselines.
These results clarify the role of geometric structure in quantum-inspired architectures and demonstrate that their benefits can be reproduced through principled geometric inductive biases implemented entirely within classical deep learning models.
\end{abstract}

\maketitle

\section{Introduction}

Over the past decade, advances in deep learning have fundamentally transformed the field of Earth Observation. Modern satellite missions such as Sentinel-2 provide high-resolution multispectral imagery with rich spectral and spatial structure, enabling fine-grained land use and land cover analysis at continental scale. Convolutional Neural Networks (CNNs) have become the dominant paradigm for extracting spatial features from such data, achieving strong performance across tasks including classification, segmentation, and change detection. 

However, multispectral and hyperspectral imagery exhibit an intrinsic high-dimensional structure: spectral bands are not independent channels but physically correlated measurements determined by material reflectance properties. Standard CNN architectures, originally designed for RGB imagery, typically treat these channels as independent Euclidean features and therefore do not explicitly incorporate the geometric or physical structure underlying spectral mixing and atmospheric transformations. This raises an important question: can architectural priors informed by geometry improve the representation and learning of multispectral satellite data \cite{NEURIPS2021}?

Recent developments \cite{Geometric_Survey} in geometric deep learning (GDL) suggest that embedding structural priors directly into neural network architectures can significantly improve both efficiency and generalization. Rather than relying solely on increasing model size, these approaches incorporate symmetries, invariances, or manifold structure into the learning process \cite{bronstein2021geometric,new_citation1}. Examples include architectures designed for non-Euclidean domains such as graphs, grids, and manifolds, where geometric constraints help restrict the hypothesis space to functions consistent with the underlying structure of the data \cite{new_citation2}. In remote sensing, such geometric priors can help capture relationships between spectral bands and spatial patterns that are not easily represented in standard convolutional architectures.

A related line of research has emerged in quantum machine learning (QML), where classical data are embedded into quantum states through unitary transformations \cite{new_citation2}. Although practical quantum hardware remains limited, the mathematical framework underlying quantum evolution provides a structured way of generating high-dimensional feature representations. In particular, unitary group actions naturally introduce symmetries and invariances that can act as inductive biases during learning. This observation has motivated a growing interest in quantum-inspired models, which aim to reproduce some of these geometric properties within classical machine learning (ML) architectures.

In the context of Earth Observation, several recent studies have explored hybrid quantum–classical models for remote sensing tasks \cite{new_citation3}. For example, Sebastianelli et al. demonstrated that circuit-based hybrid quantum neural networks can classify satellite imagery and may offer advantages in certain low-data regimes \cite{sebastianelli2021circuit, sebastianelli2022hybrid,new_citation4}. However, translating these ideas into practical quantum implementations remains challenging due to the limitations of current noisy intermediate-scale quantum (NISQ) hardware \cite{new_citation5}. In addition, increasing circuit depth to represent more complex feature maps can lead to training difficulties, such as vanishing gradients associated with barren plateaus \cite{ragone2024lie, mhiri2025constrained, new_citation7}. 

In this work, we take a different approach. Rather than relying on quantum hardware, we extract a key geometric mechanism underlying many quantum feature maps—namely the action of unitary transformations on a reference state—and incorporate it directly into a classical convolutional architecture. Building on ideas from quantum-inspired autoencoders \cite{new_citation8}, we introduce a quantum-inspired pooling operation that can be integrated into standard CNNs. This mechanism preserves some of the geometric properties exploited by quantum feature maps while remaining fully implementable on classical hardware. We demonstrate that this approach improves optimization stability and convergence in the classification of multispectral satellite imagery.

\section{Methods}

\subsection{$\SU(d)$ Unitary Pooling Layer}

We introduce a quantum-inspired pooling module that can be integrated into a classical neural network (NN). The module maps latent classical feature vectors to unitary transformations acting on a fixed quantum state, thereby embedding classical information into a Hilbert space representation. We refer to this differentiable component as the \textit{$\SU(d)$ pooling layer}.

Let $\mathbf{x} \in \R^{d^2-1}$ denote a feature vector produced by an intermediate layer of a neural network. We use the entries of $\mathbf{x}$ as coefficients of generators of the Lie algebra $\su(d)$, allowing us to construct a unitary operator. 

Let $\{\hat{G}_k\}_{k=1}^{d^2-1}$ be an orthonormal basis of $\su(d)$, such as the Gell-Mann matrices for $d=3$. We define the Hermitian generator $\hat{H}(\mathbf{x})$ and the associated unitary transformation $\hat{U}(\mathbf{x})$ via the exponential map:
\begin{equation}
    \hat{H}(\mathbf{x}) = \sum_{k=1}^{d^2-1} [\mathbf{x}]_k \hat{G}_k, 
    \qquad
    \hat{U}(\mathbf{x}) = \exp\!\left(i\,\hat{H}(\mathbf{x})\right) \in \SU(d).\label{U}
\end{equation}

To obtain a pooled representation, we apply this unitary transformation to a fixed reference state. Let $\{\ket{j}\}_{j=0}^{d-1}$ denote the computational basis of the $d$-dimensional Hilbert space $\C^d$, and consider the reference state $\ket{0}$. The resulting state is
\begin{equation}
    \ket{\psi(\mathbf{x})} = \hat{U}(\mathbf{x}) \ket{0}. \label{psi}
\end{equation}

To interface with the subsequent real-valued layers of the NN, we map the complex state vector to real coordinates by extracting its real and imaginary components. Specifically, we define
\begin{equation}
    r_j = \mathrm{Re}\!\left(\braket{j|\psi(\mathbf{x})}\right), 
    \qquad
    s_j = \mathrm{Im}\!\left(\braket{j|\psi(\mathbf{x})}\right),
    \quad j=0,\ldots,d-1 .
\end{equation}

These coefficients are assembled into a real vector $\Phi(\mathbf{x}) \in \R^{2d}$ according to
\begin{align}
    [\Phi(\mathbf{x})]_{2j} &= r_j,\nonumber\\
    [\Phi(\mathbf{x})]_{2j+1} &= s_j,
    \qquad j=0,\ldots,d-1 .\label{phi}
\end{align}

By construction, the representation satisfies $\|\Phi(\mathbf{x})\|_2 = 1$, ensuring that the pooled features lie on a compact manifold in $\R^{2d}$.

\subsection{Quotient Geometry and Dimensionality Reduction \label{quot}}

We first note that the representation $\ket{\psi(\mathbf{x})}$ in Eq.~(\ref{psi}), and consequently the vector $\Phi(\mathbf{x})$ in Eq.~(\ref{phi}), depends only on the orbit of the unitary operator $\hat{U}(\mathbf{x})$ acting on the reference state $\ket{0}$. In other words, different unitary transformations that act identically on $\ket{0}$ produce the same output representation. This construction therefore removes degrees of freedom that do not affect the state, leading to a natural reduction of the representation space.

More formally, the stabilizer subgroup of the reference state $\ket{0}$ consists of all unitary transformations that leave the state unchanged up to a global phase. As a consequence, if two input vectors $\mathbf{x}$ and $\mathbf{y}$ generate unitary operators that differ only by an element of this stabilizer group, they produce identical outputs, i.e.,
\begin{equation*}
\Phi(\mathbf{x})=\Phi(\mathbf{y}).
\end{equation*}
We refer to this effect as \textit{non-identifiability collapse}: certain directions in the parameter space of $\SU(d)$ become unobservable in the representation.

The image of $\Phi(\mathbf{x})$ can therefore be identified with the complex projective space $\CP^{d-1}$, which arises naturally as the quotient space associated with this construction. Since $\ket{\psi(\mathbf{x})}$ is a normalized quantum state, its representation is invariant under multiplication by a global phase, and therefore corresponds to a point in projective Hilbert space. More precisely, the construction can be viewed as the composition
\begin{equation*}
\mathbf{x} \;\mapsto\; \hat{U}(\mathbf{x}) \;\mapsto\; \hat{U}(\mathbf{x})\ket{0} \;\mapsto\; [\hat{U}(\mathbf{x})\ket{0}] ,
\end{equation*}
where $[\cdot]$ denotes the equivalence class under global phase, identifying vectors that differ only by multiplication with a complex scalar of unit modulus. The resulting representation therefore lives naturally on the projective Hilbert space $\CP^{d-1}$.

Although the vector $\Phi(\mathbf{x})$ is embedded in $\R^{2d}$, the representation manifold $\mathcal{M}$ has intrinsic real dimension
\begin{equation*}
\dim \mathcal{M} = 2d-2 .
\end{equation*}
For example, in the case $d=3$ used in our experiments, the effective optimization takes place on a manifold of dimension $4$, rather than in the ambient space $\R^{6}$.

This dimensionality reduction is a direct consequence of the geometric structure of the representation. In contrast to data-driven dimensionality reduction methods such as Principal Components Analysis (PCA) or autoencoders, which depend on dataset statistics, the reduction here follows from the algebraic structure of the unitary group and the choice of reference state. As a result, the learning process is effectively constrained to a lower-dimensional manifold, which can improve the conditioning of the optimization problem.

\subsection{Optimization Dynamics and Jacobian Rank Deficiency}

The geometric collapse described above has direct implications for the trainability of the NN. To analyze this effect, we consider the Jacobian $J_{\Phi}(\mathbf{x})$ of the pooling layer. Since the mapping $\Phi(\mathbf{x})$ factors through the lower-dimensional manifold $\CP^{d-1}$, the Jacobian is necessarily rank-deficient, satisfying
\begin{equation*}
\mathrm{rank}\, J_{\Phi}(\mathbf{x}) \le 2d-2 .
\end{equation*}
This rank deficiency reflects the presence of directions in the parameter space of $\SU(d)$ that do not affect the output representation. In other words, variations along the stabilizer subgroup leave $\Phi(\mathbf{x})$ unchanged and therefore lie in the kernel of $J_{\Phi}(\mathbf{x})$.

In many NN architectures, such symmetry-induced invariances can lead to optimization ambiguities, where gradients allow parameters to drift along flat directions without reducing the loss \cite{paper3, new_new_2}. In our construction, these directions are removed at the level of the representation itself. Because variations along the stabilizer subgroup are mapped to zero by the Jacobian, the optimizer updates only the directions that affect the intrinsic geometry of the representation manifold.

Consequently, the learning dynamics effectively evolve on the quotient manifold $\CP^{d-1}$ rather than in the higher-dimensional parameter space. This structural constraint reduces optimization ambiguity and is consistent with theoretical and empirical findings that symmetry-aware parameterizations can improve convergence and stability during training \cite{paper3, new_new_3}.

\section{Experiments}

We demonstrate that the theoretical benefits of quotient geometry translate into empirical performance through experiments on remote sensing data. In particular, we benchmark the proposed architecture against several classical baselines in order to isolate the effect of the unitary pooling layer.

To validate the framework we use the EuroSAT dataset, a standard benchmark for land-use and land-cover classification derived from Sentinel-2 satellite imagery \cite{helber2019eurosat}. The dataset contains 27,000 images distributed across 10 classes. In contrast to many studies that use only RGB information, we employ all 13 spectral bands available in Sentinel-2 imagery, resulting in a high-dimensional input space that is challenging for architectures optimized for three-channel data. Using all spectral bands allows the model to exploit the full spectral signatures of land-cover classes, which are often lost in RGB-only representations. The images are split into training,  and test sets ($80-20$). All models are implemented in PyTorch and trained using the Adam optimizer for 100 epochs. To assess statistical robustness, each experiment is repeated over 15 independent runs with different random seeds.

To determine a suitable dimension $d$ for the unitary pooling layer, we performed preliminary experiments with $d \in \{2,3,4,5\}$. Although this was not intended as a full hyperparameter search, it allows us to examine the trade-off between geometric capacity and optimization stability. Increasing $d$ enlarges the intrinsic representation manifold $\CP^{d-1}$, whose real dimension scales as $2d-2$. This increases expressive capacity but reduces the structural regularization induced by the quotient collapse. Empirically, we observe that very small dimensions (e.g., $d=2$) lead to underfitting, while larger values ($d \ge 4$) provide only marginal accuracy improvements at the cost of slower convergence and higher variance across runs. The choice $d=3$, corresponding to an intrinsic manifold of real dimension $4$ (i.e., $\CP^2$), consistently provides the best balance between test set accuracy, convergence speed, and stability, and is therefore used in all experiments reported below.

For the empirical evaluation we construct five architectures designed to isolate the effects of architectural priors, network depth, and parameter constraints. The input to all models consists of multispectral satellite images represented as tensors of size $13 \times 64 \times 64$, where $13$ denotes the spectral channels and the remaining dimensions correspond to the spatial layout of the image. Each channel represents reflectance values at a different wavelength band, so the input can be viewed as a stack of $13$ grayscale images of size $64 \times 64$. All models operate on the same input representation to ensure that performance differences arise solely from architectural choices.

The five architectures provide controlled comparisons between classical and quantum-inspired pooling mechanisms. Model~1 serves as the classical baseline corresponding to the shallow quantum-inspired architecture of Model~4. Both share the same strided convolutional backbone, which progressively reduces spatial resolution before flattening the representation into a vector. The difference lies in the pooling stage: Model~1 employs a standard dense bottleneck, while Model~4 replaces this component with the unitary pooling layer. This pairing therefore isolates the effect of the geometric transformation while keeping the rest of the architecture unchanged.

Models~2 and~3 provide deeper classical baselines corresponding to the deep hybrid architecture introduced in Model~5. These architectures use a deeper convolutional feature extractor composed of convolution--pooling blocks that gradually reduce spatial resolution while increasing the number of feature channels. Model~2 adopts the same dimensional bottleneck structure as the quantum-inspired model, constraining the representation through layers of sizes $d^2-1$ and $2d$ in order to mirror the dimensional structure induced by the unitary pooling map. Model~3 removes this constraint and instead uses wider dense layers, providing a purely classical benchmark with greater representational flexibility. Model~5 combines the same deep convolutional backbone with the quantum-inspired pooling mechanism, enabling a direct comparison with the deep classical architectures.

\begin{figure}[t]
    \centering
    \includegraphics[width=\textwidth]{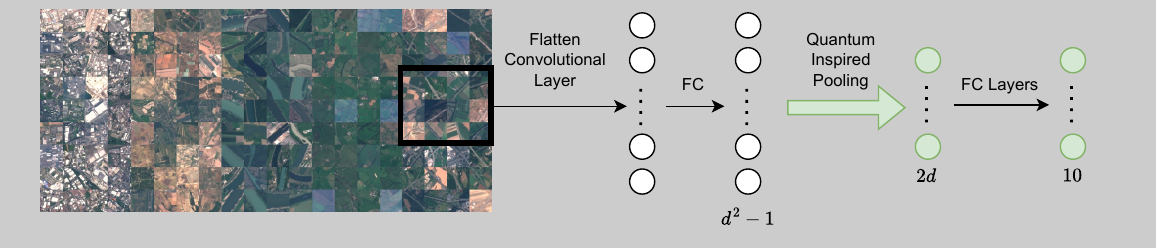}
    \caption{\textbf{Architecture of the quantum-inspired pooling CNN.} The pipeline begins with a standard CNN backbone extracting spatial features from the satellite imagery (left). These features are flattened and projected into the Lie algebra dimension ($d^2-1$) via a fully connected (FC) layer. The quantum-inspired pooling layer then maps these values to the special unitary group $SU(d)$, applies the unitary action to a reference state, and outputs a normalized vector of size $2d$. Finally, standard FC layers classify the representation into the target classes.}
    \label{fig:architecture}
\end{figure}

Taken together, these five models form a controlled experimental framework: Model~1 versus Model~4 compares classical and quantum-inspired pooling under identical shallow feature extractors, while Models~2 and~3 serve as deep classical baselines against which the hybrid Model~5 can be evaluated. This design allows us to distinguish whether performance differences arise from the unitary pooling mechanism itself, from increased architectural depth, or from differences in parameter capacity. All convolutional layers use $3\times3$ kernels with padding $=1$, unless otherwise stated.

\begin{itemize}

\item \textbf{Model 1: Shallow classical  CNN.} 
A direct classical counterpart to Model 4, using the same strided convolutional backbone:
\[
\begin{aligned}
13 \times 64 \times 64 
&\xrightarrow{\text{Conv}(32, s=2)} 32 \times 32 \times 32 \xrightarrow{\text{Conv}(64, s=2)} 64 \times 16 \times 16 \\
&\xrightarrow{\text{Conv}(64, s=2)} 64 \times 8 \times 8 \xrightarrow{\text{Flatten}} 4096
\end{aligned}
\]

The representation is then projected through a classical bottleneck mimicking the dimensional structure of the quantum model:
\[
4096 \to (d^2-1) \to 2d \to 64 \to \text{10},
\]
without imposing any geometric constraint.

\item \textbf{Model 2: Deep  classical CNN with restricted bottleneck.}
A deeper architecture employing convolution–maxpool blocks:
\[
13 \times 64 \times 64
\;\xrightarrow{\text{Conv}(40)}\;
40 \times 64 \times 64
\;\xrightarrow{\text{Pool}}\;
40 \times 32 \times 32
\]
\[
\xrightarrow{\text{Conv}(64)}
64 \times 32 \times 32
\xrightarrow{\text{Pool}}
64 \times 16 \times 16
\]
\[
\xrightarrow{\text{Conv}(80)}
80 \times 16 \times 16
\xrightarrow{\text{Pool}}
80 \times 8 \times 8
\xrightarrow{\text{Flatten}}
5120.
\]

The flattened representation is forced through the same dimensional bottleneck structure as the quantum model:
\[
5120 \to (d^2-1) \to 2d \to 64 \to \text{10},
\]
testing whether the imposed dimensional restriction alone explains the performance gains.

\item \textbf{Model 3: Deep  classical CNN (without bottleneck).}
Uses the identical deep convolutional backbone as Model 2:
\[
13 \times 64 \times 64
\to
80 \times 8 \times 8
\to
5120,
\]

but replaces the geometric bottleneck with wide dense layers:
\[
5120 \to 256 \to 128 \to 64 \to \text{10}.
\]

This model represents the upper bound of classical representational capacity without structural constraints.

\item \textbf{Model 4: Shallow  CNN with quantum-inspired pooling.}
A shallow CNN backbone with strided convolutions:
\[
13 \times 64 \times 64
\to
32 \times 32 \times 32
\to
64 \times 16 \times 16
\to
64 \times 8 \times 8
\to
4096,
\]

followed by the unitary pooling map:
\[
4096
\to
\mathfrak{su}(d) \sim (d^2-1)
\xrightarrow{\exp(iH)}
SU(d)
\curvearrowright |0\rangle
\to
\mathbb{CP}^{d-1}
\hookrightarrow
2d,
\]
and final classification layers:
\[
2d \to 64 \to \text{10}.
\]

\item \textbf{Model 5: Deep  CNN with quantum-inspired pooling.}
A deep convolutional architecture with two classical spatial downsampling stages:
\[
13 \times 64 \times 64
\;\xrightarrow{\text{Conv}(40)}\;
40 \times 64 \times 64
\;\xrightarrow{\text{Pool}}\;
40 \times 32 \times 32
\]
\[
\xrightarrow{\text{Conv}(64)}
64 \times 32 \times 32
\xrightarrow{\text{Pool}}
64 \times 16 \times 16
\]
\[
\xrightarrow{\text{Conv}(80)}
80 \times 16 \times 16
\xrightarrow{\text{Flatten}}
20480.
\]

The high-dimensional semantic representation is then mapped to the Lie algebra:
\[
20480
\to
\mathfrak{su}(d) \sim (d^2-1)
\xrightarrow{\exp(iH)}
SU(d)
\curvearrowright |0\rangle
\to
\mathbb{CP}^{d-1}
\hookrightarrow
2d,
\]

followed by the classifier:
\[
2d \to 64 \to \text{10}.
\]

This hybrid architecture isolates the effect of quotient geometry within a deep feature extractor.

\end{itemize}

The architecture of our proposed method is visualized in Figure \ref{fig:architecture}, which illustrates the transition from high-dimensional spatial features to the compact unitary manifold. All models were trained for 100 epochs across 15 independent runs to ensure statistical significance. The numerical performance metrics are summarized in Table \ref{tab:results}.

\begin{table}[htbp]
\centering
\footnotesize
\setlength{\tabcolsep}{5pt}
\renewcommand{\arraystretch}{1.0}
\caption{Performance on the 10-class EuroSAT dataset. Best test accuracy, convergence speed (epochs to 80\% and 90\%), peak epoch, and number of trainable parameters.}
\label{tab:results}
\begin{tabular}{lccccc}
\toprule
\textbf{Model} & \textbf{\# Params} & \textbf{Max Test Acc (\%)} & \textbf{80\%} & \textbf{90\%} & \textbf{Peak Epoch} \\
\midrule
Model 1 (Shallow Classical) & 134,088 & 80.81 & 39.13 & 78.53 & 85.53 \\
Model 2 (Deep Classical w/ bottleneck)   & 198,492 & 92.96 & 23.93 & 60.53 & 96.53 \\
Model 3 (Deep Classical w/o bottleneck) & 1,426,762 & 94.60 & 7.80 & 22.00 & 90.87 \\
Model 4 (Shallow Quantum Inspired)   & 93,074 & 93.97 & 7.73 & 19.73 & 71.80 \\
\textbf{Model 5 (Deep Quantum Inspired)} & 238,930 & \textbf{94.78} & \textbf{4.53} & \textbf{10.07} & \textbf{71.27} \\
\bottomrule
\end{tabular}
\end{table}

\begin{figure}[htbp]
    \centering
    \includegraphics[width=0.9\textwidth]{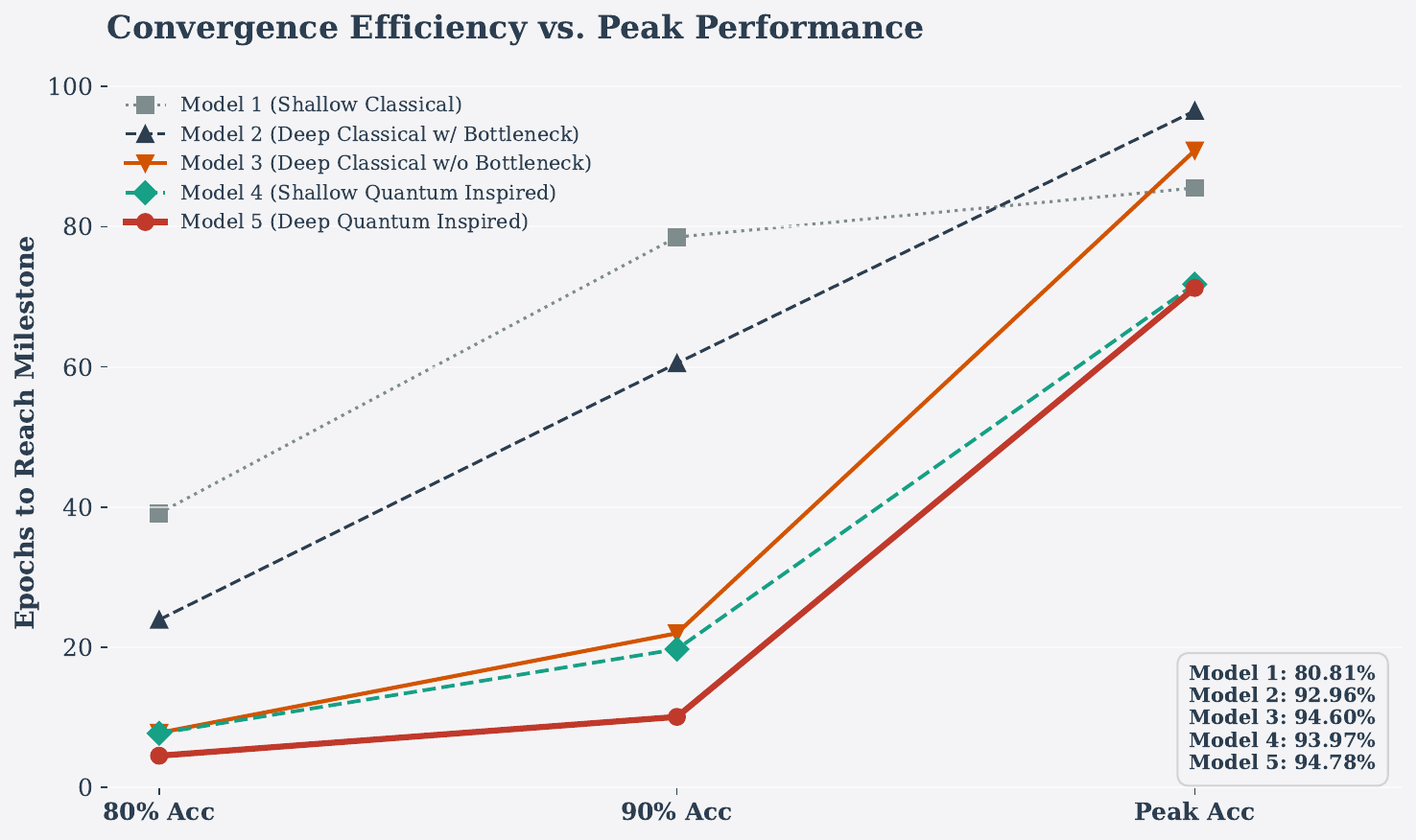}
    \caption{\textbf{Convergence Efficiency vs. Peak Performance.} The number of epochs required to reach 80\% and 90\% test accuracy milestones, alongside the epoch of peak performance. The Deep Hybrid model (red circle, Model 5) demonstrates superior convergence velocity, reaching 90\% accuracy in approximately 10 epochs—twice as fast as the strongest classical baseline (Model 4). Lower values on the y-axis indicate faster learning.}
    \label{fig:convergence}
\end{figure}

The Deep Hybrid model (Model 5) achieves a competitive test set accuracy of $94.78\%$, marginally outperforming the unrestricted Deep unrestricted classical benchmark (Model 3, $94.60\%$) despite operating under strict geometric constraints. This demonstrates that the $SU(d)$ pooling layer acts as a highly effective regularizer that does not compromise representational capacity. Furthermore, the performance gap between the Deep Restricted model (Model 2, $92.96\%$) and Model 5 confirms that the advantage stems from the quotient geometry itself, rather than the mere presence of a dimensional bottleneck.

Crucially, the most pronounced advantage lies in convergence velocity and optimization stability, as illustrated in Figure \ref{fig:convergence}. The Deep Hybrid model reaches 90\% accuracy in just $10.07$ epochs, compared to $22.00$ epochs for the strongest classical baseline (Model 3)—a twofold improvement in learning speed. This rapid convergence is consistent with our theoretical analysis of Jacobian rank deficiency, which suggests the optimizer avoids ``flat'' directions in the landscape corresponding to the stabilizer subgroup. The shallow quantum model (Model 4) also demonstrates remarkable efficiency, reaching 90\% accuracy faster ($19.73$ epochs) than the unrestricted deep baseline, further validating the efficiency of the geometric prior. Importantly, the per-epoch wall-clock time is effectively identical across models (122.6 seconds for the Deep Unrestricted model and 123.9 seconds for the Deep Hybrid model), confirming that the observed gains correspond to a genuine reduction in total training time rather than an artifact of differing computational costs per epoch.

While deep data re-uploading circuits in QML often face a trade-off where increasing depth leads to barren plateaus \cite{ragone2024lie}, our method avoids this by employing a fixed, shallow geometric bottleneck. The results confirm that this ``quantum'' nonlinearity, when applied deterministically, preserves gradient magnitude and significantly accelerates training compared to both restricted and unrestricted classical architectures.

\section{Discussion}

This section discusses the geometric interpretation of the proposed pooling layer and its implications for representation learning and optimization.

To better understand the mechanism of the $\SU(d)$ pooling layer, it is useful to compare it with established classical approaches. PCA, for example, assumes that data lie approximately on a linear subspace and seeks directions that maximize variance. When the underlying data manifold is curved—as is often the case for complex spectral signatures—such linear projections may distort geometric relationships. In contrast, the $\SU(d)$ pooling layer implements a nonlinear transformation that maps feature vectors to the projective space $\CP^{d-1}$. This mapping preserves the normalized structure of the representation while reducing dimensionality through the quotient geometry described in Section~\ref{quot}.

The proposed mechanism also differs fundamentally from standard pooling operations such as max or average pooling. Classical pooling acts as a local scalar operation that aggregates values within a neighborhood, often discarding correlations between feature channels. By contrast, $\SU(d)$ pooling mixes the input features through a unitary transformation whose parameters are determined by the feature vector itself. The resulting state representation preserves global structure while compressing the feature space through the geometry of the underlying manifold.

We emphasize that the proposed method does not claim a fundamental computational advantage over classical NNs in the complexity-theoretic sense. Rather, its contribution is conceptual: it identifies a concrete geometric mechanism—quotient-induced non-identifiability collapse—that explains why certain quantum-inspired constructions can exhibit favorable training behavior even when implemented on classical hardware. From this perspective, the $\SU(d)$ pooling layer can be interpreted as a structured geometric regularizer, similar in spirit to group-equivariant convolutions or manifold-constrained representations in geometric deep learning.

Finally, the normalization inherent in the construction ensures that the output of the $\SU(d)$ layer lies on a compact manifold with fixed norm. In contrast to unconstrained Euclidean activations, this geometric constraint naturally bounds the representation space and can stabilize the optimization dynamics. This interpretation is consistent with recent work on Density Quantum Neural Networks, which suggests that structured state-space representations can improve trainability \cite{coyle2025training}. In our setting, the quotient geometry of $\CP^{d-1}$ plays a similar role by removing redundant degrees of freedom and constraining the learning dynamics to the intrinsic representation manifold. More broadly, this perspective highlights how geometric structures originally arising in quantum feature maps can inspire new architectural components for classical deep learning models.

\section{Conclusion}

In this work, we investigated how geometric ideas originating from quantum feature maps can be incorporated into classical deep learning architectures for remote sensing. By introducing an $\SU(d)$ pooling layer that applies unitary transformations to latent feature vectors, we obtained a nonlinear representation that maps data onto the projective space $\CP^{d-1}$.

Our analysis shows that this construction naturally removes redundant degrees of freedom through the quotient geometry of the representation. As a result, the effective optimization takes place on a lower-dimensional manifold, which can help stabilize training. The numerical experiments on the EuroSAT dataset indicate that incorporating this pooling mechanism leads to improved convergence behavior and reduced variance compared to standard pooling layers.

More broadly, our results suggest that certain benefits often associated with quantum-inspired models may stem from their underlying geometric structure rather than from quantum hardware itself. From this perspective, the $\SU(d)$ pooling layer provides a simple way to introduce such geometric priors within fully classical NN architectures. Future work will investigate how similar geometric constructions can be incorporated into larger architectures and evaluated on a wider range of remote sensing tasks.

\section*{Acknowledgments}
This work was supported by the European Union’s Horizon Europe research and innovation program under grant agreement No.101092766 (ALLEGRO Project).

\section*{Availability}
Data availability: The EuroSAT dataset is publicly available at \url{https://github.com/phelber/EuroSAT} and via Zenodo. Code availability: The code corresponding to the $\SU(d)$ pooling implementation is available upon reasonable request.

\bibliographystyle{plain}
\bibliography{references}

\end{document}